# Viewpoint – Utilizing Data from Prevalent Technologies for Atmospheric Research

Noam David[1]

*Institute of Industrial Science, The University of Tokyo, Tokyo, Japan*

Over the past decade, the Internet of Things (IoT) and smart devices have become increasingly common as part of the technological infrastructure that surrounds us. The flow of data generated by these systems is characterized by enormous granularity, availability and coverage. As a result, new opportunities are being opened to utilize the newly available information for various needs and, in particular, for atmospheric research. If we consider the data generated by these means, we may notice that many produce measurements with high environmental value. To name some examples - surveillance cameras that operate in the visible light spectrum are positioned in a vast number of locations. Previous works have shown that they can be used for monitoring the temporal patterns of fine atmospheric particulate matter (Wong et al., 2007). Lab experiments indicate a direct link between the speed of movement of car wipers, and the rainfall intensity. Hence, advanced vehicles that store these data can, in essence, be used as moving rain gauges (Rabiei et al., 2013). Kawamura et al (2017) have revealed a novel technique for monitoring atmospheric humidity using terrestrial broadcasting waves, based on propagation delays due to water vapor. Data shared as open source from social networks have been found to be potentially effective in improving automatic weather observations. Indeed, for the most part, the initial weather observation is conducted

---

[1]Correspondence to: Dr. Noam David

E mail: nd363@cornell.edu

automatically by dedicated sensors, however, some weather conditions are still better detected by the human eye. On the other hand, millions of 'human observers' around the world use applications such as Twitter, which allows them to report publicly on subjects that are relevant to them, and in particular on weather phenomena (Cox and Plale, 2011). As was recently reviewed by Price et al (2018), in 2020 there will be more than 20 billion smartphones carried by the public worldwide. These mobile devices are equipped with sensors that can be used for environmental monitoring on a multisource basis. Recent works indicate the ability to obtain atmospheric temperature information for the urban canopy layer (Overeem et al., 2013a), to measure atmospheric pressure (Mass and Madaus, 2014), or to study atmospheric tides (Price et al., 2018). Additional studies point to the potential of using any camera enabled smart mobile device to monitor air quality (Pan et al., 2017).

Given the comprehensive coverage of the new 'virtual sensors' from all land locations across the whole globe, this low cost solution introduces a wide range of possibilities which previously could not be offered through existing technologies.

A key example of utilizing exiting data sources for atmospheric monitoring is the use of measurements acquired by commercial microwave links (CMLs) that comprise the infrastructure for wireless data transport between cellular base stations. This technology has essentially become a valuable weather monitoring tool over the past decade (e.g. Overeem et al., 2013b; Alpert et al., 2016; Gosset et al., 2016; David and Gao, 2017).

Weather phenomena and atmospheric conditions interfere with the electromagnetic waves transmitted by CMLs. Thus, these networks provide, in essence, an already existing environmental monitoring facility. The various works done in the field indicate the ability to detect and map rainfall (e.g. Messer et al., 2006; Berne and Uijlenhoet, 2007; Chwala et al., 2012; Overeem et al., 2013b; Fencl et al., 2015), monitor fog, atmospheric moisture and dew (e.g. Chwala et al., 2014; David, 2014; David et al., 2016). Recent research revealed the ability of these radio links to identify atmospheric refractive index variations and to indirectly detect air pollution conditions (David and Gao, 2016; 2018).

Indeed, the new data available from the various means (smartphones, social networks, etc) and particularly from CMLs can provide observations with considerable spatial coverage and with minimal cost. However, the accuracy of each 'sensor' is lower than that of a dedicated instrument. This being the case, is it possible to produce significant information compared to that derived from specialized tools?

It can be assumed that the 'virtual sensors' are not a substitute for conventional monitoring means, whenever those exist in the field. The correct approach, then, is to consider the newly available sources of data as complementary measures to dedicated measurements and as a substitute during the many cases in which conventional monitoring tools are not available. However, the data acquired by prevalent technologies, even when taken by itself, often holds enormous potential.

In order to demonstrate the added value which lies in IoT data and prevalent technologies, let us focus on CMLs as an example of such a system.

Aatmospheric moisture is more poorly characterized than wind or even precipitation due to the difficulty in observing the humidity field. Therefore, questions such as the magnitude of small-scale variability of moisture in the boundary layer, and its effect on convection initiation are still unanswered (Weckwerth et al., 2004). As a result, prediction of convective precipitation, on the storm scale, is limited. However, for significantly improving convection initiation measurements, one will need moisture measurements at meso-γ resolution with accuracies of up to 1 gr/m$^3$ (Fabry, 2006). Notably, such type of observations can be acquired using CMLs (David et al., 2019). High resolution precipitation distribution maps can be generated using CMLs, therefore the relationship between pollutant wash-off and rainfall provides an opportunity to potentially acquire important spatial information about air quality, as discussed in recent research (David and Gao, 2016). Moreover, Liquid Water Content (LWC) constitutes a major parameter in fog research. Fog LWC changes in space, altitude, and over time, and is dependent on surface and atmospheric conditions (Gultepe et al., 2007). However, conventional sensors for acquiring LWC estimates are limited in the spatial range they can cover and in their availability. It has been shown that CMLs are able to provide fog LWC estimates across large spatial regions where dedicated sensors are nonexistent. Indeed, the availability of various spectral channels from satellites provides the possibility to observe clouds, aerosols, Earth surface and fog - in particular (Lensky and Rosenfeld, 2008; Michael et al., 2018). However, CMLs have also been found to have potential advantages for detecting fog under challenging conditions where satellite retrievals are limited, e.g when high altitude clouds cover the fog as observed from the satellite vantage point (David, 2018). Alternatively, the ability to monitor rainfall in areas where radars suffer from clutter effects (Goldshtein et al., 2009) or are blocked by complex topography (David, 2014), has also been demonstrated.

The possibilities for monitoring atmospheric phenomena utilizing the new observational powers are many, the available information vast, and the cost minimal, since the 'opportunistic sensors' are already deployed in the field. As a result, this mean of monitoring the environment is becoming advantageous for atmospheric research. Notably, the newly available 'virtual sensors' open the possibility of assimilation of their measurements into high-resolution numerical prediction models which could lead to improvements in the forecasting capabilities that exist today. In the applicative aspect, this novel approach could lay the groundwork for developing new early warning systems against natural hazards, generating a range of products required for a wide range of fields and, thus, the overall potential contribution to public health and safety may be invaluable.

**Acknowledgments**

I wish to gratefully acknowledge Associate Professor Dr. Yoshihide Sekimoto and his research team for the fruitful discussions and for hosting me in their laboratory as a Visiting Scientist at the Institute of Industrial Science of the University of Tokyo, Japan during 2018-2019.